\documentclass[journal]{IEEEtran}
\usepackage{amsmath,amsfonts}
\usepackage{amssymb}
\usepackage{hyperref}
\usepackage[utf8]{inputenc}
\usepackage{algorithm}

\usepackage{times}
\usepackage{array}
\usepackage[caption=false]{subfig}
\usepackage{textcomp}
\usepackage{stfloats}
\usepackage{fix-cm}
\usepackage{authblk}
\usepackage{bbm}
\usepackage{bm}
\usepackage{nicefrac}
\usepackage{microtype}
\usepackage{mathtools}
\usepackage{hyperref}
\usepackage{cleveref}
\usepackage{colortbl}
\usepackage{ragged2e}
\usepackage{url}
\usepackage{verbatim}
\usepackage[pdftex]{graphicx}
\usepackage{cite}

\usepackage{tikz}
\usetikzlibrary{shapes,arrows.meta,positioning,shapes.geometric,fit,calc,arrows.meta,shadows,backgrounds}

\definecolor{userblue}{RGB}{70, 130, 180}   
\definecolor{sysgray}{RGB}{105, 105, 105}   
\definecolor{alertred}{RGB}{178, 34, 34}    
\definecolor{procColor}{RGB}{230, 245, 255}   
\definecolor{procBorder}{RGB}{70, 130, 180}   
\definecolor{ioColor}{RGB}{255, 245, 230}     
\definecolor{ioBorder}{RGB}{210, 105, 30}     
\definecolor{termColor}{RGB}{230, 255, 230}   
\definecolor{termBorder}{RGB}{34, 139, 34}    
\definecolor{simColor}{RGB}{235, 240, 245}     
\definecolor{agentColor}{RGB}{255, 248, 240}   
\definecolor{blockBorder}{RGB}{60, 60, 60}     
\definecolor{highlight}{RGB}{0, 110, 180}      
\definecolor{acColor}{RGB}{240, 248, 255}    
\definecolor{dcColor}{RGB}{255, 245, 238}    
\definecolor{ctrlColor}{RGB}{245, 255, 250}  
\definecolor{lineColor}{RGB}{60, 60, 60}     
\definecolor{powerLine}{RGB}{20, 20, 20}     

\usepackage{todonotes}
\usepackage{listings}
\usepackage{booktabs}
\usepackage[table]{xcolor}
\usepackage{pifont}
\usepackage{xspace}
\usepackage{tabularx}
\usepackage{algpseudocode}
\usepackage[inkscapelatex=false]{svg}

\usepackage{ragged2e}

\usepackage{array}

\newcommand{\pythonname}[1]{\texttt{#1}\xspace}
\newcommand{\SystemComponent}{\pythonname{Sys\-tem\-Com\-po\-nent}}
\newcommand{\Inverter}{\pythonname{In\-ver\-ter}}
\newcommand{\PowerSource}{\pythonname{Po\-wer\-Source}}
\newcommand{\Clock}{\pythonname{Clock}}
\newcommand{\Simulator}{\pythonname{Si\-mu\-la\-tor}}
\newcommand{\Battery}{\pythonname{Bat\-te\-ry}}
\newcommand{\Grid}{\pythonname{Grid}}
\newcommand{\Load}{\pythonname{Load}}
\newcommand{\step}{\pythonname{step}}
\newcommand{\stepticks}{\pythonname{step\_\-ticks}}
\newcommand{\abs}[1]{\lvert {#1} \rvert}

\begin{document}


\title{Bridging Natural Language and Microgrid Dynamics: A Context-Aware Simulator and Dataset}


\author{Tinko Sebastian Bartels,
Ruixiang~Wu, Xinyu~Lu, Yikai~Lu, Fanzeng~Xia, Haoxiang~Yang, Yue~Chen, Tongxin~Li
\thanks{T. Bartels, R. Wu, X. Lu, Y. Lu, F. Xia, H. Yang, and T. Li are with the School of Data Science, The Chinese University of Hong Kong-Shenzhen, Shenzhen, China.}
\thanks{Y. Chen is with the Department of Mechanical and Automation Engineering, The Chinese University of Hong Kong, Hong Kong, China.}
\thanks{This work has been submitted to the IEEE for possible publication. Copyright may be transferred without notice, after which this version may no longer be accessible.}
}
\markboth{arXiv Preprint.}%
{Bartels \MakeLowercase{\textit{et al.}}: The OpenCEM Simulator for Context-Rich Energy Analysis}


\maketitle

\begin{abstract}
Addressing the critical need for intelligent, context-aware energy management in renewable systems, we introduce the \textbf{OpenCEM Simulator and Dataset}: the first open-source digital twin explicitly designed to integrate rich, unstructured contextual information with quantitative renewable energy dynamics. Traditional energy management relies heavily on numerical time series, thereby neglecting the significant predictive power embedded in human-generated context (e.g., event schedules, system logs, user intentions). OpenCEM bridges this gap by offering a unique platform comprising both a meticulously aligned, language-rich dataset from a real-world PV-and-battery microgrid installation and a modular simulator capable of natively processing this multi-modal context.
The OpenCEM Simulator provides a high-fidelity environment for developing and validating novel control algorithms and prediction models, particularly those leveraging Large Language Models. We detail its component-based architecture, hybrid data-driven and physics-based modelling capabilities, and demonstrate its utility through practical examples, including context-aware load forecasting and the implementation of online optimal battery charging control strategies. By making this platform publicly available, OpenCEM aims to accelerate research into the next generation of intelligent, sustainable, and truly context-aware energy systems.
\end{abstract}


\section{Introduction}
\IEEEPARstart{D}{ecarbonizing}  the power grid hinges on managing intermittent renewables like solar and batteries, making accurate energy forecasting a critical challenge~\cite{impram2020challenges}. While traditional energy research has focused on numerical time series, this approach fails to capture the why behind energy fluctuations, ignoring the rich predictive intelligence found in human-generated context.

\newcommand{\cmark}{\ding{51}}
\newcommand{\xmark}{\ding{55}}
\definecolor{Green}{RGB}{40, 167, 69}
\definecolor{Red}{RGB}{220, 53, 69}
\newcommand{\greentick}{{\color{Green}\cmark}}
\newcommand{\redcross}{{\color{Red}\xmark}}

In practice, the true behavior of an energy system is often dictated by real-world events that are not directly captured in standard sensor readings. For instance, a simple natural language statement like, \textit{Tomorrow I will run a CPU-intensive, multi-core numeric robustness test for a day,} contains more predictive power about future energy load than hours of historical time-series data. Such contextual information—found in event calendars, maintenance logs, user announcements, and even social media—provides direct insight into future energy needs but exists in unstructured, multi-modal formats. This presents the fundamental challenge of \textit{modeling and simulating the impact of this rich, qualitative context on the quantitative dynamics of a renewable energy system.}

This challenge has become critically important with the advent of Large Language Models (LLMs) and Foundation Models (FMs) capable of understanding and reasoning over natural language~\cite{dong2024surveyincontextlearning,instructmpc}. For the first time, we have the computational tools to interpret this contextual data and translate it into actionable intelligence for energy management. Yet, a significant gap prevents progress, i.e., the lack of appropriate fundamental infrastructure for contextual decision-making in real physical renewable energy systems. Existing microgrid data sets~\cite{dataset_microgrid_mesa_del_sol,dataset_microgrid_nanogreen_japan, dataset_microgrid_rye} provide fine-grained numerical time series but lack detailed natural language context information. To develop and validate new context-aware models, researchers in the power and energy society need two things that do not currently exist:
\begin{enumerate}
    \item \textbf{A language-rich dataset} that provides meticulously aligned time series of electrical measurements (load, generation, battery state) alongside the corresponding unstructured, real-world context that influenced them.
    \item \textbf{A simulation environment} capable of natively processing this contextual information, allowing researchers to go beyond historical replays and to test how hypothetical scenarios or new control strategies would perform.
\end{enumerate}

This challenge is fundamental to tasks such as context-aware battery scheduling within microgrids, where the objective is to leverage cheap grid power to anticipate future high loads, minimizing costs while ensuring reliability~\cite{fernandez2019power}. Success in these dynamic environments requires predictive models that bridge the gap between simple numerical time series and real-world events. While numerical weather forecasts provide one layer of physical insight, the most critical predictive signals often reside in unstructured natural language. Such context ranges from structured event schedules to purely textual data, such as event reports, user announcements, or system logs indicating  computationally intensive software compilations. Translating this diverse, natural language context to directly inform physical microgrid dynamics is the primary motivation for integrating LLMs and FMs into energy management~\cite{dong2024surveyincontextlearning, moeini2025surveyincontextreinforcementlearning}.

Training and evaluating these advanced models, however, demands a new testing infrastructure. Researchers require testbeds that  synchronize traditional power metering time series with rich, multimodal contextual data. Such a resource is currently unavailable, as existing datasets are either too narrow in scope or proprietary.
Existing state-of-the-art simulators, while powerful for modelling physical dynamics, are fundamentally context-agnostic (see Table~\ref{tab:simulator_features_abbr}). They cannot process a textual event description to simulate its effect on the grid. This tooling and data gap creates a major bottleneck, thus hindering the development of the next generation of intelligent, context-aware energy systems.

To bridge this gap, in this paper we present the \textbf{OpenCEM Simulator}. Based on the Open In-Context Energy Management Platform (OpenCEM)~\cite{opencem_demo}, it collects energy generation, battery level, and load time series from an on-campus PV installation. This system utilizes batteries to supply power to a university room equipped with a research workstation, air conditioner, and other varying loads brought in when the room is in use. The OpenCEM Simulator is an open-source digital twin that consists of both a language-rich dataset and a context-aware simulator for renewable energy research. Grounded in a real-world, instrumented PV-and-battery installation,
OpenCEM is the first platform designed to explicitly model the interplay between qualitative context and physical power flows. We demonstrate the simulator's capabilities and the unique insights provided by its dataset through practical examples, providing a foundational tool for the research community.

\textbf{Our Contributions.} To bridge this critical gap between qualitative context and quantitative energy modelling, this paper makes the following primary contributions:

\begin{itemize}
    \item \textbf{An Open-Source Context-Aware Simulator:} We present the design and implementation of the OpenCEM simulator, a modular, open-source digital twin. It is the first simulation framework designed to natively integrate and model the impact of unstructured, time-stamped contextual data on physical power system dynamics.

    \item \textbf{A Unique Language-Rich Dataset:} We introduce and release a unique public OpenCEM dataset containing synchronized electrical time series and multi-source, multi-modal context from a live renewable energy system, including textual and semi-structured event data. This dataset provides the first real-world corpus for training, fine-tuning, and validating context-aware renewable energy models.

    \item \textbf{Testbed Validation:} We validate the simulator against real-world data from our testbed and present use cases that offer foundational insights into the system. Through these examples, we demonstrate the strong, quantifiable correlation between textual context and energy profiles, showcasing the simulator's utility for future context-rich energy analysis.
    
\end{itemize}

The simulator source code, dataset database, as well as notebooks with usage examples and the code to produce the graphs in this work are available at 
\begin{center}
\href{https://github.com/OpenCEM-platform/opencem_simulator}
{\textit{https://github.com/OpenCEM-platform/opencem\_simulator}}.
\end{center}

\begin{figure*}\centering\includegraphics[alt={Figure of the components of the OpenCEM platform including the database, algorithmic controller/simulator, the website which serves as a frontend to students and researchers, context data sources such as news, event calendar, and internet, the power system installation including PV array, battery, campus building, and grid connection.},width=0.85\textwidth]{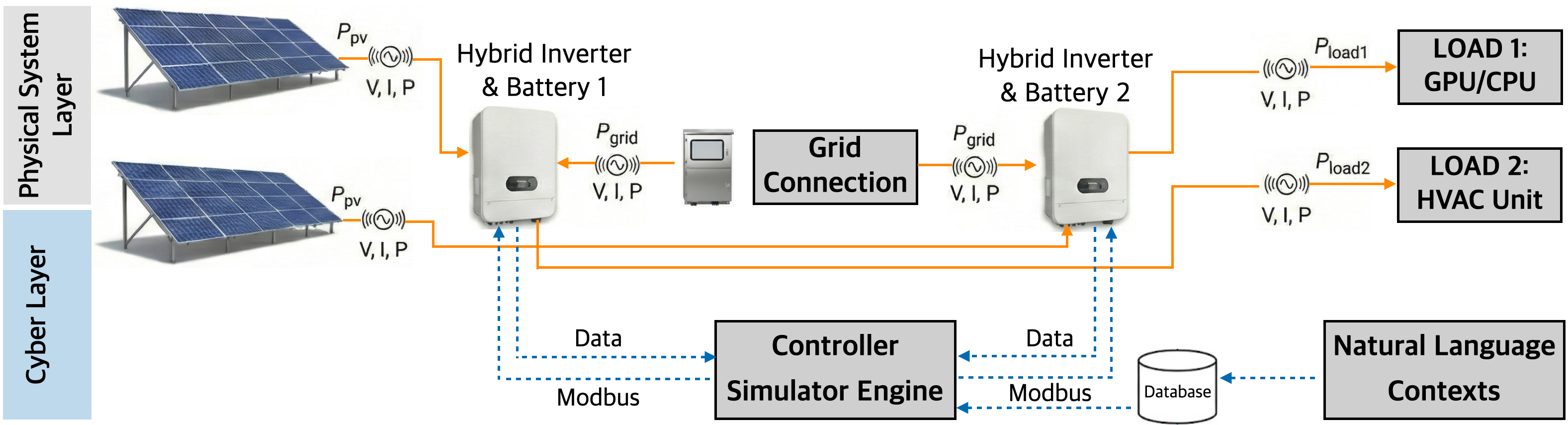}
  \caption{\textbf{High-level Architecture of the OpenCEM Platform.} The framework is divided into two domains: (Top) The \textbf{Physical System Layer} comprises two independent microgrid subsystems, where PV arrays and hybrid inverters with battery storage power distinct loads—a research workstation (GPU/CPU) and an HVAC unit. (Bottom) The \textbf{Cyber Layer} interfaces with the hardware via Modbus to log electrical measurements ($V$: Voltage, $I$: Current, $P$: Power) and integrates this quantitative data with unstructured \textbf{Natural Language Contexts} in a central database to drive the Controller Simulator Engine.}
  \label{fig:framework}
\end{figure*}
\begin{table}[t] 
\centering
\footnotesize 
\setlength{\tabcolsep}{4pt} 
\renewcommand{\arraystretch}{1.1}
\caption{Feature Comparison of Open Source Microgrid Simulators.}
\label{tab:simulator_features_abbr}
\begin{tabular}{l cccccc}
\toprule
\textbf{Simulator} & \textbf{PV} & \textbf{Batt.} & \textbf{Inv.} & \textbf{Grid} & \textbf{Load} & \textbf{Ctx.} \\
\cmidrule(r){1-1} \cmidrule(lr){2-7} 
MATPOWER~\cite{5491276} & \redcross & \redcross & \redcross & \greentick & \greentick & \redcross \\
ANDES~\cite{cui2020hybridsymbolicnumericframeworkpower} & \redcross & \redcross & \redcross & \greentick & \greentick & \redcross \\
PS.Dynamics~\cite{lara2024powersimulationsdynamicsjlopensource} & \redcross & \redcross & \greentick & \greentick & \greentick & \redcross \\
pvlib~\cite{Anderson2023} & \greentick & \redcross & \greentick & \redcross & \redcross & \redcross \\
ACN-Sim~\cite{8909765} & \redcross & \redcross & \redcross & \greentick & \greentick & \redcross \\
EV2Gym~\cite{Orfanoudakis_2025} & \redcross & \redcross & \redcross & \greentick & \greentick & \redcross \\
pycity\_sch.~\cite{SCHWARZ2021100839} & \greentick & \greentick & \redcross & \greentick & \greentick & \redcross \\
\midrule
\rowcolor{gray!20} 
\textbf{OpenCEM} (Ours) & \greentick & \greentick & \greentick & \greentick & \greentick & \greentick \\
\bottomrule
\end{tabular}
\vspace{-10pt}
\end{table}

\textbf{Related Works.}
Open-source power system simulators have a long history of supporting research and development in smart grid technologies. The landscape of these tools has evolved significantly from classic power flow optimization problems to the integration of renewable energy and electric vehicles~(EVs).

MATPOWER~\cite{5491276}, released in 2011, provides an accessible, open-source MATLAB package for steady-state power system simulation, providing tools for optimal power flow~(OPF) analysis. Its key feature is an extensive OPF architecture, which allows the users to customize variables, costs, and linear constraints of the simulated power system. The package uses this framework to implement advanced features in real power systems like piecewise linear cost, dispatchable loads, and generator capability curves. Building on this, ANDES~\cite{cui2020hybridsymbolicnumericframeworkpower} revisited the challenge of modelling power system dynamics via differential-algebraic-equations. Its symbolic layer allows researchers to define dynamic power system components using equation strings and built-in blocks like transfer functions and limiters, from which it automatically generates computationally efficient code and Jacobians for the system modelling. More recently, PowerSimulationsDynamics.jl~\cite{lara2024powersimulationsdynamicsjlopensource} was developed in Julia to handle the dynamic response of modern power systems with high penetrations of Inverter-Based Resources~(IBRs). An essential capability is its support for both Quasi-Static Phasor and Electro-Magnetic Transient simulations. It features a modular structure for IBRs and uses Automatic Differentiation to compute Jacobians.

The challenge of accurately modelling the power system dynamics, driven by the global trend of decarbonization, stimulated the occurrence of a new batch of specialized simulators. To handle the special features of solar energy, pvlib~\cite{7750303} provides a set of weather-to-energy generation functions to model the conversion chain from solar irradiance to AC power generation, and it has been updated recently to include features like bifacial modules and losses from  soiling and snow~\cite{Anderson2023}. In addition, the rapid rise of EVs necessitated new tools like ACN-Sim~\cite{8909765}, a data-driven simulator designed to evaluate online scheduling algorithms by running them on a real-world charging station data ACN-Data~\cite{10.1145/3307772.3328313}. Afterwards, EV2Gym~\cite{Orfanoudakis_2025} provides a standardized Open AI Gym~\cite{brockman2016openaigym} environment for benchmarking smart charging algorithms, with a focus on Vehicle-to-Grid scenario. It offers a comprehensive simulation at the power transformer level by incorporating realistic models for EV battery degradation, diverse EV types, and inflexible loads.

To address the operational challenge of urban energy system, pycity\_scheduling~\cite{SCHWARZ2021100839} provides a framework for developing and assessing optimization-based power scheduling algorithms, with a special design for multi-energy systems at the city level. This special design enables the co-optimization of coupled electricity and thermal sectors. The framework's primary function is to solve the day-ahead power dispatch problem to achieve system-level objectives like cost minimization or peak-shaving.

While these simulators provide useful models for modelling the physical components of the grid with features summarized in Table~\ref{tab:simulator_features_abbr}, none of the existing simulators are equipped to process contextual information, such as weather forecasts, policy changes, or social events, when provided in a natural language format. This restricts their ability to capture how real-world conditions dynamically influence power generation and consumption.

\vspace{-6pt}
\section{System Framework and Physical Testbed}

\begin{figure}[t!] 
    \centering
    \subfloat[\rm Aerial view of the rooftop PV installation (Array 1 \& 2).]{
\includegraphics[width=0.75\columnwidth]{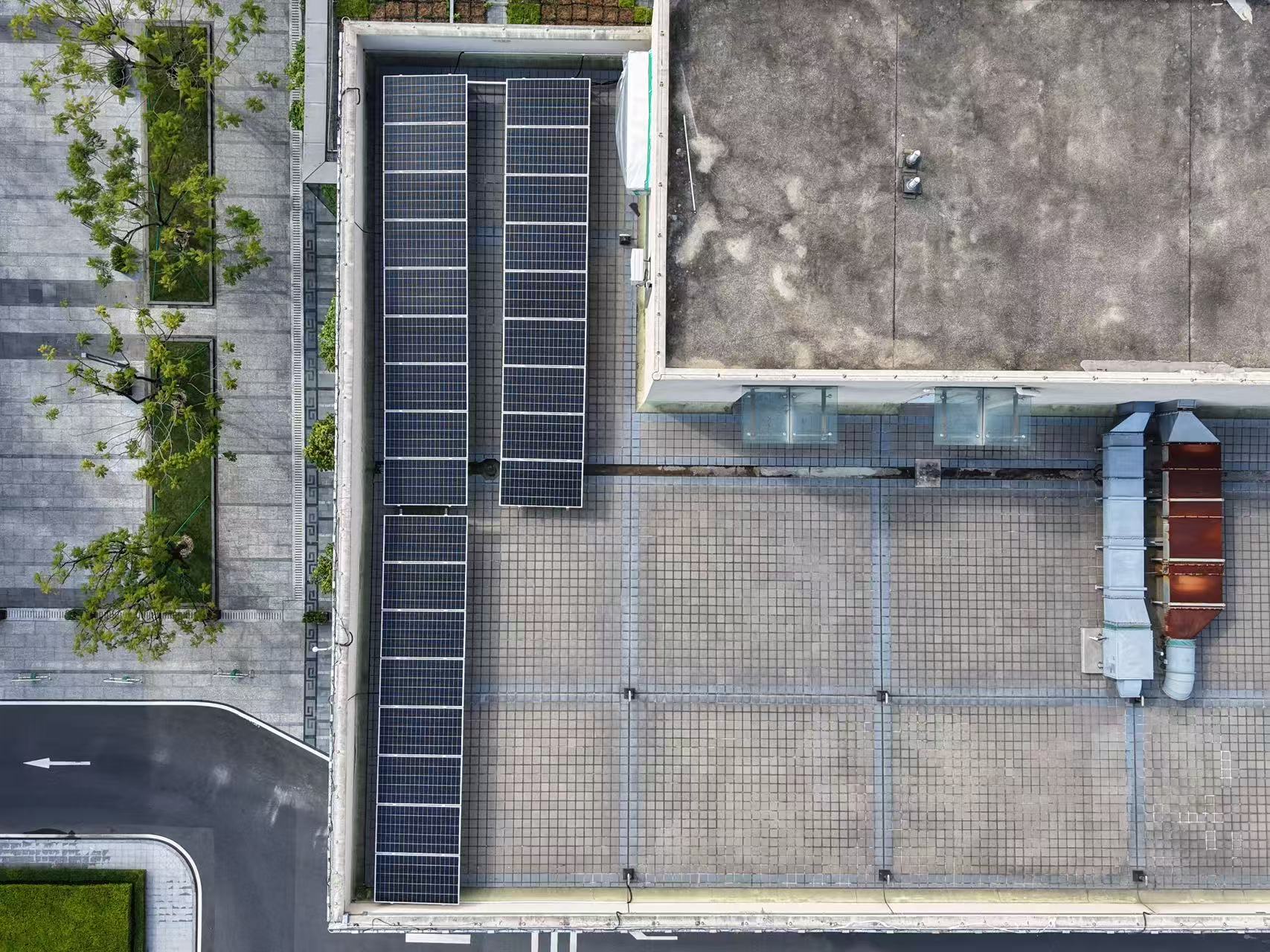}
    \label{fig:pv_aerial}}
    
    \subfloat[\rm Wall-mounted hybrid inverters and battery storage.]{
\includegraphics[width=0.75\columnwidth]{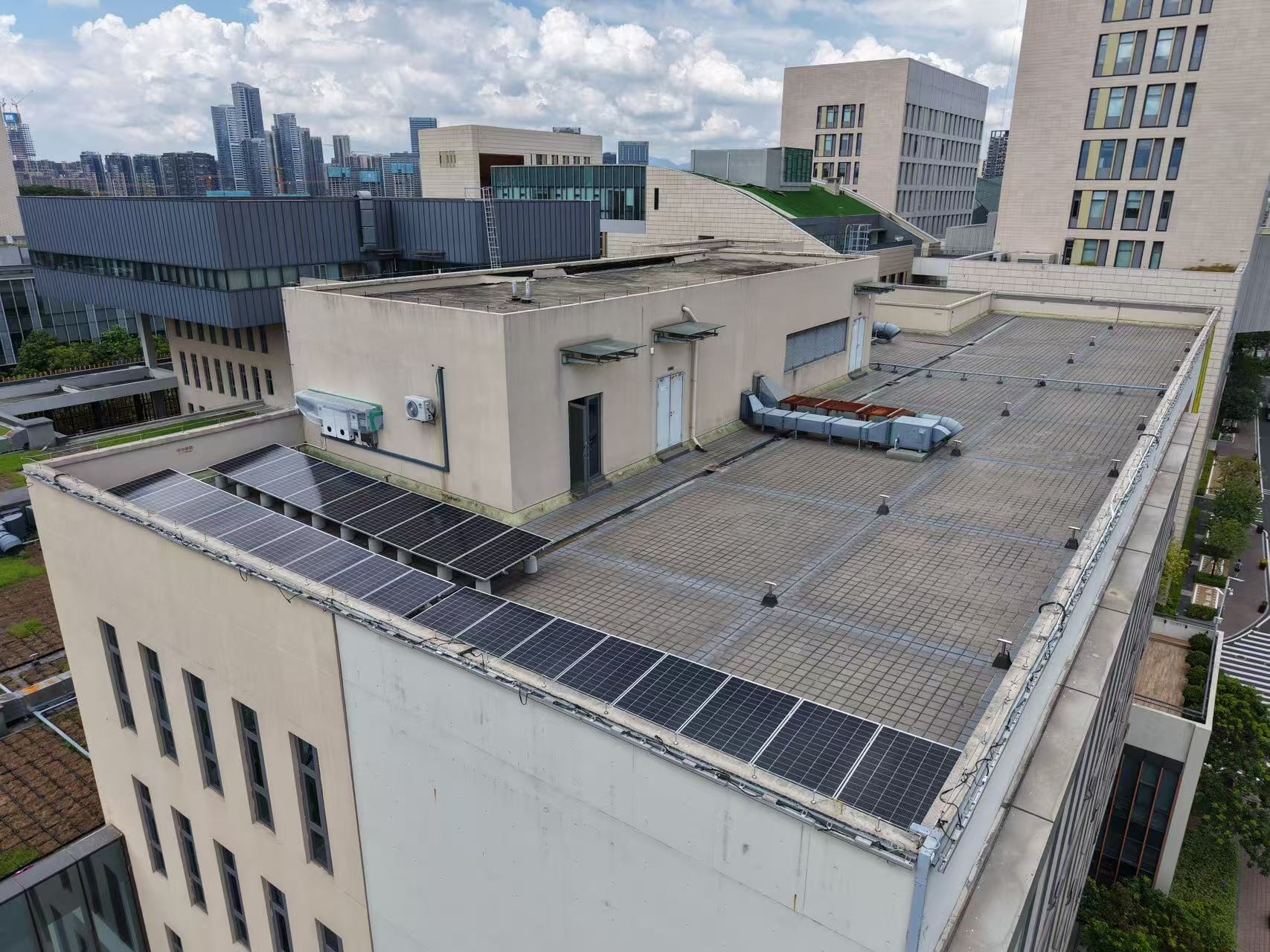}
        \label{fig:inverters}}
\caption{\rm \textbf{Implementation of the OpenCEM Physical Layer.} The system consists of (a) two distinct PV arrays on the facility roof and (b) the corresponding hybrid inverter control units.}
\label{fig:physical_system_combined}
\vspace{-11pt}
\end{figure}

The OpenCEM framework operates as a high-fidelity digital twin, anchored by a fully instrumented real-world testbed as shown in Figure~\ref{fig:physical_system_combined}. This section details our simulator's architecture, which are decomposed into two main aspects. In Section~\ref{sec:physical installation}, we introduce the simulator's physical layer: an physical micro power system that provides real-world data for the simulator. Subsequently, in Section~\ref{sec:modular,CBA}, we describe the simulator's cyber layer, detailing its modular, component-based architecture which mirrors the physical system described in Section~\ref{sec:physical installation}. The special design of the framework provides an extensible environment that allows researchers to conveniently develop and test novel control strategies.

\vspace{-6pt}
\subsection{On-Campus Microgrid} \label{sec:physical installation}
The physical installation that the simulator models is a microgrid located on our university campus. It provides the high-fidelity time-series data used in the simulator's data-driven mode. The system includes two PV arrays (with 26 panels), each with its own inverter and lithium-ion battery pack, powering dynamic loads such as research workstations and an air conditioner.


Both inverters are instrumented to capture detailed electrical measurements at two-minute intervals, forming the core of the dataset described in Section~\ref{sec:dataset}. Instantaneous electric readings, as well as statistics recorded by the inverters are read from the inverters through a USB-to-RS485 connection which allows communication via a Modbus protocol. Modbus is a client/server communication protocol and a de-facto standard in many industrial applications, see~\cite{modbus2012} for more details. In our application it serves for both the reading of measurements and the sending of control signals.

In addition, context information for decision-making is recorded from event 
announcements, scraped web data, the university schedule, workstation logs, 
and user-generated input. The dataset will be open to the research community to enable research 
in this critical area.



\begin{figure*}
  \centering\includegraphics[alt={Representative series of power drawn from grid over two days of usage for inverter 2, which powers an air conditioner.},width=0.95\textwidth]{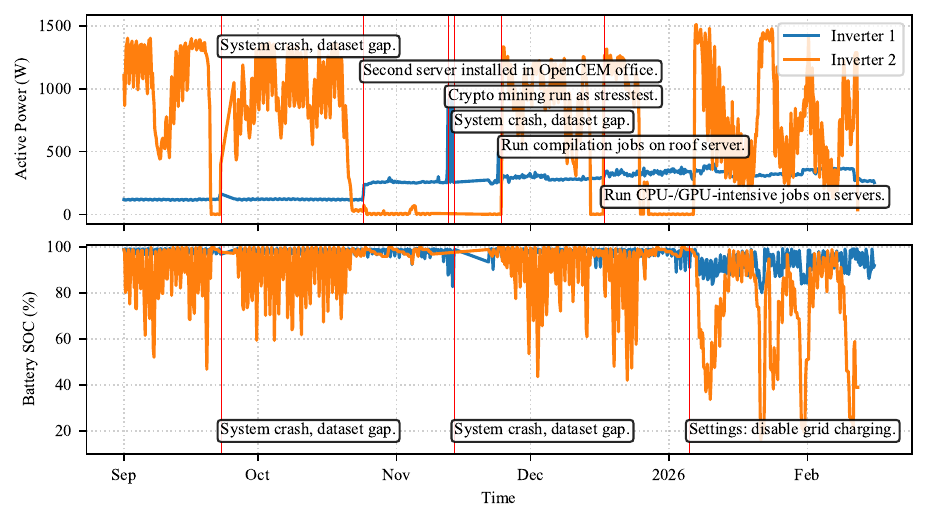}\\
  \vspace{-15pt}
  \caption{\textbf{Power Load and Battery SOC} over time with context annotations. Example Time Series with Electrical Measurements. \textnormal{Power drawn by load, and battery SOC for inverter 1, which powers two servers, and inverter 2, which powers an air conditioner.}}
  \label{fig:example-time-series}
\end{figure*}

\vspace{-5pt}
\subsection{Modular, Component-Based Architecture} \label{sec:modular,CBA}
The simulator is built on a modular, object-oriented architecture that mirrors the physical system, as shown in Figure~\ref{fig:framework}. The core components are represented as distinct classes: a \texttt{PowerSource} (PV array), a \texttt{Battery} (BESS), a \texttt{Load}, a \texttt{Grid} connection, an \texttt{Inverter} that manages power flow between them, and a \pythonname{Context} class that exposes textual context information about future events when such information becomes available.

This component-based design offers two significant advantages. First, it allows users to easily assemble and configure different system setups by combining components. Second, it promotes extensibility, enabling users to implement and test their own models—for instance, a more sophisticated battery degradation model or a novel inverter control strategy—by simply creating a new class that adheres to the base component's interface. This modularity is crucial for isolating variables and systematically evaluating the performance of specific algorithms. A detailed overview of the component APIs is provided in Section~\ref{sec:api}.



\section{The OpenCEM Dataset}
\label{sec:dataset}

\begin{figure}[h]
  \includegraphics[width=0.45\textwidth]{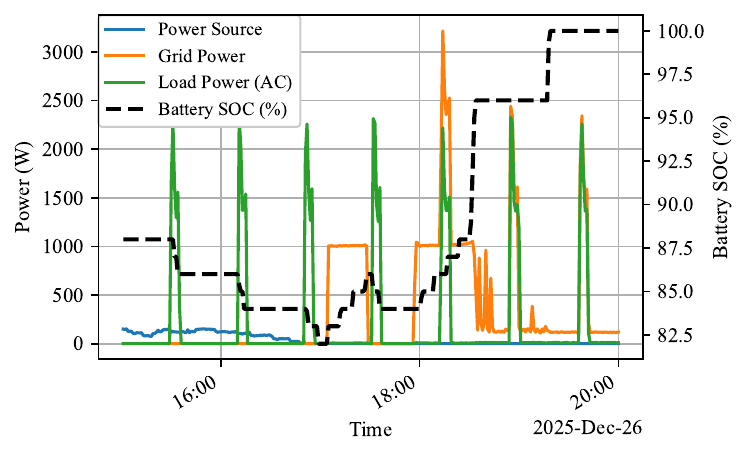}\\
  \vspace{-10pt}
  \caption{\textbf{Representative Power Flows and Battery SOC} over one day. Example Time Series with Electrical Measurements. \textnormal{Representative series (sampled on 2025.12.26) of power drawn from grid, power generation, battery SOC, and power demand of the load over five hours of usage for inverter 2, which powers an air conditioner.}}
  \label{fig:example-time-series-day}
\end{figure}

The microgrid's inverter exposes a number of physical measurements, such as
readings of voltage, current, power (apparent and active for AC circuits) at all relevant contacts, readings from the battery's BMS, operating parameters,
diagnostic readings, metadata, settings, and more.
The raw dataset includes all parameters, while for the simulator, a selection 
was made of the most relevant readings for ease of use.
In Figure~\ref{fig:example-time-series-day} and Figure~\ref{fig:example-time-series}, we show charts of measurements taken from the system over a representative day and over several months respectively.


The initial dataset covers the period from July 2025 to January 2026, with plans for regular updates in the project repository.
Measurements are taken roughly every two minutes, with the limiting factor 
being the Modbus interface.

\subsection{Electrical System Time Series}\label{subsec:electrical-system-time-series}

\definecolor{row_highlight}{gray}{1}

In the following, we list the most important electrical measurements, as provided in the dataset by component. For a more complete list of recorded values, see the linked repository.

\subsubsection{Battery}
The continuous electrical measurements of the inverter at the battery include 
voltage (V), current (A, signed depending on charging or discharging), 
state of charge (SOC, \%), and charging power (W).
Additionally, values from the BMS are recorded for the requested charging 
voltage (V), discharge voltage limit (V), requested charging current (A),
and requested discharge current (A).
The inverter also exposes aggregate statistics of total charging, and 
discharging energy (kWh), over the current day, last week, and total lifetime 
of the system.
Relevant settings, whose values are recorded in the dataset, include over- and
under-voltage alarm thresholds (V), SOC thresholds (\%), for forcing the 
charging from grid power or stopping any charging, upper and lower SOC limits, 
and voltage and timing controls for the CV phase of charging.

All values returned by the BMS are included in the dataset, to allow validation of a range of 
models from simple linear approaches based on nominal voltage, capacity and 
SOC, to more fine-grained physical simulations that take different charging 
phases, and full charging and discharging voltage curves into account.

\subsubsection{PV Array}

For the PV array, we report continuous measurements of voltage (V), current (A), power (W), no controllable settings, and statistics of generated power (kWh) over time windows of one day, one week, and the total lifetime of 
the system.

\subsubsection{Load}
The system powers AC loads, for which voltage (V), current (A), frequency (Hz), active power (W), apparent power (VA), and the percentage of the output limit (\%) are provided for each time step in the dataset.
Statistics include total power consumed (kWh) in total and directly from the current day, last seven days, and the total lifetime of the system.
Controllable parameters include the output target voltage (V) and frequency 
(Hz), which are fixed for our application to 230 V and 50 Hz, respectively.

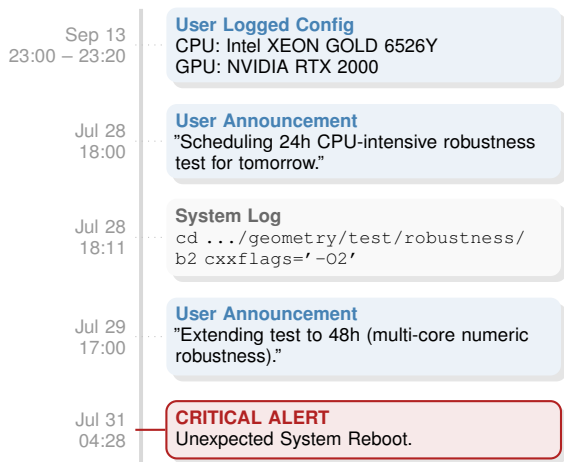
\begin{figure}[htbp]
\centering
\begin{tikzpicture}[scale=0.8,
    node distance=0cm,
    every node/.style={font=\sffamily\scriptsize},
    time/.style={
        text width=2cm, 
        align=right, 
        color=gray!80,
        font=\sffamily\scriptsize
    },
    dot/.style={
        circle, 
        fill=white, 
        draw=gray, 
        thick, 
        inner sep=0pt, 
        minimum size=5pt
    },
    eventbox/.style={
        rectangle, 
        rounded corners, 
        draw=none, 
        fill=gray!10, 
        text width=5.0cm, 
        align=left, 
        inner sep=3.5pt,
        drop shadow={opacity=0.2}
    }
]

\draw[draw=gray!30, line width=1.5pt] (-0.9,0.6) -- (-0.9,-7);

\node[time] (t1) at (-2.4, 0) {Sep 13\\23:00 -- 23:20};
\node[dot, draw=userblue, fill=userblue] (d1) at (0, 0) {};
\node[eventbox, fill=userblue!10] (e1) at (2.8, 0) {
    \textbf{\textcolor{userblue}{User Logged Config}}\\
    CPU: Intel XEON GOLD 6526Y\\
    GPU: NVIDIA RTX 2000
};

\node[time] (t2) at (-2.4, -1.6) {Jul 28\\18:00};
\node[dot, draw=userblue, fill=userblue] (d2) at (0, -1.6) {};
\node[eventbox, fill=userblue!10] (e2) at (2.8, -1.6) {
    \textbf{\textcolor{userblue}{User Announcement}}\\
    "Scheduling 24h CPU-intensive robustness test for tomorrow."
};

\node[time] (t3) at (-2.4, -3.2) {Jul 28\\18:11};
\node[dot, draw=sysgray, fill=sysgray] (d3) at (0, -3.2) {};
\node[eventbox, fill=gray!5] (e3) at (2.8, -3.2) {
    \textbf{\textcolor{sysgray}{System Log}}\\
    \texttt{cd .../geometry/test/robustness/}\\
    \texttt{b2 cxxflags='-O2'}
};

\node[time] (t4) at (-2.4, -4.85) {Jul 29\\17:00};
\node[dot, draw=userblue, fill=userblue] (d4) at (0, -4.85) {};
\node[eventbox, fill=userblue!10] (e4) at (2.8, -4.85) {
    \textbf{\textcolor{userblue}{User Announcement}}\\
    "Extending test to 48h (multi-core numeric robustness)."
};

\node[time] (t5) at (-2.4, -6.4) {Jul 31\\04:28};
\node[dot, draw=alertred, fill=alertred] (d5) at (0, -6.4) {};
\node[eventbox, fill=alertred!10, draw=alertred, thick] (e5) at (2.8, -6.4) {
    \textbf{\textcolor{alertred}{CRITICAL ALERT}}\\
    Unexpected System Reboot.
};

\draw[gray!50, dotted] (t1) -- (e1.west);
\draw[gray!50, dotted] (t2) -- (e2.west);
\draw[gray!50, dotted] (t3) -- (e3.west);
\draw[gray!50, dotted] (t4) -- (e4.west);
\draw[alertred, thick] (t5) -- (e5.west);

\end{tikzpicture}
\caption{\textbf{Examples of Events and Contexts.} The dataset captures both high-level user intents (Source: Team) and low-level system events (Source: Log). \textit{(Note: Context text is reproduced verbatim from dataset records).}}
\label{fig:context_timeline}
\end{figure}

\subsubsection{Grid Connection}
The project's installation is connected to the power grid one-way, so that additional power can be bought, when needed, but no selling of electricity 
is possible.
We provide measurements of grid voltage (V), current (A, signed in principle but always positive for our use case), frequency (Hz), apparent power (VA),
and active power (W).
Controllable settings include minimum and maximum limits for voltage, frequency, total current, and current used for battery charging
at the grid connection.
Statistics are recorded for energy consumed from the grid for battery charging and for the whole system (kWh) for the current day, the last seven days, and in 
total.

\subsubsection{Inverter}
The inverter exposes a wide range of internal measurements, status information, 
and controllable settings, whose full extent cannot be documented here for 
space reasons.
Controllable settings of particular interest for control algorithms include 
scheduled battery charging plans (time of day), and the inverter's policy 
for prioritizing power sources to cover the load's demand (categorical). 

\vspace{-10pt}

\subsection{Context Data}

The most significant innovation of the OpenCEM simulator is its native support for contextual information. The energy consumption and generation in modern power systems are heavily influenced by external factors like weather, scheduled events, and human behavior, which are often described in unstructured natural language. As highlighted in Table~\ref{tab:simulator_features_abbr}, no other open-source simulator is equipped to handle such information.

OpenCEM addresses this gap by including a dedicated \texttt{Context} component. This component processes and injects time-stamped contextual data—such as user-submitted plans ("running a GPU-intensive job overnight") or automated logs—into the simulation loop. By making context a first-class citizen, the simulator provides a unique testbed for models that leverage Large Language Models (LLMs) or other AI techniques to achieve more intelligent and predictive energy management.

Promoting the use of context information in power system control is a major goal of the OpenCEM project.
Relevant context for power systems is multi-modal and can include structured 
and unstructured data from different sources, where each type of context 
information can be relevant for one or more quantities that need to be 
predicted, such as load, power generation, and electricity prices.

Examples of structured context information include weather forecasts in e.g., JSON format as returned by a web provider, which affects both power generation 
and load prediction, price market data, e.g., from futures markets, which are indicators of dynamic price development, structured calendar information like 
room bookings, which is an indicator of future load, etc.

Examples of unstructured context data include full-text weather 
forecasts for load and power generation prediction, as well as natural 
language scenario descriptions by systems operators or users, or log lines of 
workstations for load prediction.

The initial dataset bundled with the simulator includes natural language 
context records with metadata about when they start and stop applying
(timestamps), when they were recorded (timestamp), to which inverter they apply
(categorical), and natural language value field.
The context records include context from both user inputs and automatically 
extracted workstation log lines. A sample of context records is given in Figure~\ref{fig:context_timeline}.


\section{The OpenCEM Simulator}
\label{sec:api}

\begin{figure}[t]
\centering
\begin{tikzpicture}[
    font=\sffamily\scriptsize,
    >={Latex[length=2mm, width=2mm]},
    node distance=0.6cm and 0.8cm, 
    block/.style={
        draw=lineColor, 
        thick, 
        fill=white, 
        rectangle, 
        rounded corners=2pt, 
        minimum height=0.9cm, 
        minimum width=1.8cm, 
        align=center,
        inner sep=3pt,
        drop shadow={opacity=0.1, shadow xshift=0.3ex, shadow yshift=-0.3ex}
    },
    inverter/.style={
        block,
        minimum height=1.4cm,
        minimum width=3.2cm,
        font=\bfseries\scriptsize,
        fill=white
    },
    power/.style={ 
        draw=powerLine, 
        line width=1.2pt
    },
    data/.style={ 
        draw=lineColor, 
        thick, 
        dashed, 
        ->
    },
    lbl/.style={
        font=\tiny\sffamily,
        color=lineColor,
        inner sep=1.5pt,
        fill=white,
        fill opacity=0.9,
        text opacity=1
    }
]


    \node[inverter] (inv) {Hybrid Inverter\\ \tiny (Central Logic)};

    \node[block, fill=ctrlColor] (clock) at ($(inv.north west) + (-0.2, 1.2)$) {\textbf{Clock}\\ \tiny (Time)};
    \node[block, fill=ctrlColor] (context) at ($(inv.north east) + (0.2, 1.2)$) {\textbf{Context}\\ \tiny (Logs/Events)};

    \node[block, fill=dcColor, left=0.8cm of inv] (batt) {\textbf{Battery}\\ \tiny (Storage)};
    \node[block, fill=dcColor, above=0.3cm of batt] (pv) {\textbf{PV Array}\\ \tiny (Generation)};

    \node[block, fill=acColor, right=0.8cm of inv] (grid) {\textbf{Grid}\\ \tiny (Import)};
    \node[block, fill=acColor, below=0.3cm of grid] (load) {\textbf{Load}\\ \tiny (Demand)};


    \draw[data] (clock.south) -- node[lbl, left] {Sync} (inv.130);
    \draw[data] (context.south) -- node[lbl, right] {Events} (inv.50);

    \draw[power, ->] (pv.east) -- (inv.170) node[lbl, pos=0.5, above] {DC};
    \draw[power, <->] (batt.east) -- (inv.190) node[lbl, pos=0.5, below] {DC};

    \draw[power, <-] (inv.10) -- (grid.west) node[lbl, pos=0.5, above] {AC};
    \draw[power, ->] (inv.350) -- (load.west) node[lbl, pos=0.5, below] {AC};

    
    \node[anchor=south, font=\bfseries\tiny, color=orange!70!black] at (pv.north) {DC Domain};
    \node[anchor=south, font=\bfseries\tiny, color=blue!70!black] at (grid.north) {AC Domain};

    \begin{scope}[on background layer]
        \node[draw=lineColor!40, dashed, fill=gray!5, rounded corners=6pt, fit=(clock) (context) (batt) (load) (grid) (pv), inner sep=5pt] (sim_frame) {};
        \node[anchor=south west, font=\bfseries\scriptsize, color=gray!80!black] at (sim_frame.south west) {Simulator Environment};
    \end{scope}

\end{tikzpicture}
\vspace{-15pt}
\caption{\textbf{System Component Topology.} The architecture segregates the \textbf{DC Domain} (PV, Battery) from the \textbf{AC Domain} (Grid, Load). The central \textbf{Hybrid Inverter} manages bi-directional power conversion, guided by data streams from the \textbf{Context} and \textbf{Clock} modules.}
\label{fig:simulator_topology}
\end{figure}
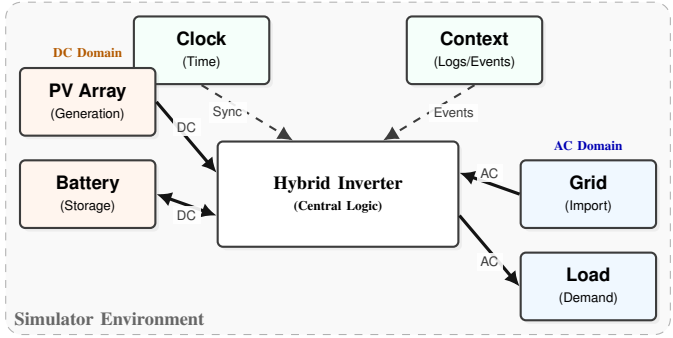

\subsection{Abstract API Overview}
The API of the simulator is designed to be modular, extensible, and generic, 
allowing simulation with combinations of different models for each component of the system, and generic, while enforcing a minimal common interface that 
each component model must implement to guarantee that basic electrical 
constraints can be verified, and all time series of interest can be computed.
In the following, we document this baseline and give motivation for the requirements that it enforces on model implementations.

\subsubsection{SystemComponent}
\SystemComponent is an abstract base class (ABC) for the different component interfaces that establishes common conventions for the simulator loop.
It specifies an abstract \step method, which takes an integer \stepticks, denoted by $\Delta t$,
and optionally more arguments which differ between different subclasses,
and returns the component's result of advancing by \stepticks steps of a shared
time resolution, i.e.:
\[
\SystemComponent.\step: \left(\Delta t, \cdot \right) \mapsto \left(\cdot\right).
\]

\subsubsection{PowerSource}
The ABC \PowerSource implements \SystemComponent, and models 
any independent DC power source, such as a solar panel array.
Its \step method returns a \pythonname{Po\-wer\-Source\-Step\-Re\-sult} dataclass object that 
contains at least the output voltage in V, current in A, and power in W at the end of the current timestep:
\begin{multline}
\nonumber
\PowerSource.\step: \\
\left(\Delta t,\cdot \right) \mapsto \left(U_{\pythonname{PS},t+\Delta t}, I_{\pythonname{PS},t+\Delta t}, P_{\pythonname{PS},t+\Delta t}, \cdot\right).
\end{multline}
The inclusion of the power field, though seemingly redundant in DC systems, is motivated by the substantial discrepancies observed between the recorded power and the product of voltage and current in the dataset.
Because it has no inherent two-way relationship with other system components,
its \step method specifies no additional mandatory arguments, but implementors
of the \PowerSource interface may specify additional arguments, e.g.,
a weather prediction.

\subsubsection{Grid}
The ABC \Grid implements \SystemComponent, and models a one-way AC grid connection that allows a connected inverter to draw power from the grid but not
sell power back.
Its \step method takes, besides \stepticks, a \pythonname{GridStepInput}, which must at least specify the requested apparent power demand in VA and active (real)
power in W.
The returned \pythonname{GridStepResult} returns the actually delivered apparent power $S$ in
VA and active power $P$ in W:
\[
\Grid.\step:\left(\Delta t,P_{\pythonname{G},\text{req},t},S_{\pythonname{G},\text{req},t},\cdot\right)\mapsto\left(P_{\pythonname{G},\text{del},t},S_{\pythonname{G},\text{del},t},\cdot\right).
\]
No voltage values are passed or retrieved because they are fixed at 230 V at 
the installation site.
Reactive power $Q$ in var is not included because it is not measured 
independently in the real system that is being modelled. 
It can be inferred from the usual relationship
$
    \abs{S} = \sqrt{P^2 + Q^2}.
$

\subsubsection{Load}
The ABC \Load implements \SystemComponent, and models an AC load supplied with
electric power at a target voltage of 230 V and frequency of 50 Hz by the 
inverter.
Its \step method takes no mandatory additional arguments besides \stepticks.
It returns a \pythonname{LoadStepResult} with its requested active power in W and apparent
power in VA, as measured at the contact to the inverter:
\begin{align*}
    \Load.\step:\left(\Delta t,\cdot\right)\mapsto\left(P_{\pythonname{L},\text{req},t},S_{\pythonname{G},\text{req},t},\cdot\right).
\end{align*}
As above, reactive power may be inferred but is not returned directly.

\subsubsection{Battery}
The ABC \Battery implements \SystemComponent, and models a DC battery, 
connected to and controlled by the inverter.
Its step method takes a \pythonname{BatteryStepInput}, which specifies the current mode 
out of the categories charging, discharging, and idle, and a current in A.
No voltage is specified because it is not controlled externally, but depends on 
the state of the battery.
It returns a \pythonname{BatteryStepResult} instance containing the SOC (out of $\left[0,1\right]$), the signed change in charge in C (negative in case of 
charging), and the signed change energy in J:
\begin{multline}    
\Battery.\step:\\
\left(\Delta t,\pythonname{Mode}_t,I_{\pythonname{B},t},\cdot\right)\mapsto\left( \text{SOC}_t,U_{\pythonname{B},t},\Delta E_{\pythonname{B},t},\Delta C_{t},\cdot \right).
\end{multline}

\subsubsection{Inverter}
The ABC \Inverter implements \SystemComponent and models a DC-AC converter connecting a DC \PowerSource and \Battery to a \Load, with the option to import power from the \Grid.
Its \step method takes \stepticks and an \pythonname{InverterStepInput} containing the previous step results. It returns an \pythonname{InverterStepResult} with:
\begin{itemize}
    \item The next \pythonname{BatteryStepInput} and \pythonname{GridStepInput};
    \item The power drawn from the generator (which may be less than available capacity, e.g., if the battery is full):
    \begin{multline}
\Inverter.\step:\\
\left(\Delta t, U_{\pythonname{PS},t},\ldots,C_{t},\cdot\right)\mapsto\left(P_{\pythonname{G},\text{req},t},\ldots,I_{\pythonname{B},t},P'_{\pythonname{PS},t},\cdot\right).
\end{multline}
\end{itemize}

This component is responsible for prioritizing between drawing power from the 
\Battery, \PowerSource, and \Grid, and meet the (active and apparent) power 
demands of the \Load at each time step, i.e.
\[
\begin{aligned}
P_{\pythonname L,\text{req},t} & \leq P'_{\pythonname{PS},t}+P_{\pythonname{G},\text{del},t}+\Delta E_{\pythonname{B},t}\cdot \Delta t^{-1}, \\
P'_{\pythonname{PS},t} & \leq P_{\pythonname{PS},t},\\
S_{L,\text{req},t} & \leq S_{\pythonname{G},\text{del},t},
\end{aligned}
\]
and its interface is intended to be
implemented by simulator users to test control algorithms.

\subsubsection{Context}
The ABC \pythonname{Context} implements \SystemComponent, and returns at each step a set of future \pythonname{ContextRecords} that were created before the time of the current step and apply to a time interval in the present or future with respect to the current step $t$:
\[
\pythonname{Context}.\step: \left(\Delta t\right) \mapsto \{ \left(t_{\text{recorded},1}, t_{\text{begin}, 1}, t_{\text{end}, 1}, \ast \right), \ldots \},
\]
where $\ast$ denotes a JSON object that may contain a natural language event description, as well as structured metadata, depending on the underlying context event. For each returned context record it holds that $t_\text{recorded} \leq t$ and $t_\text{end}>t$.

\subsection{Utility Classes}

\subsubsection{Clock}
The \Clock class is an immutable data class used to synchronize the current time $t$,
time resolution, and time steps $\Delta t$ among all system components.
It advances time on an integral ns scale internally and provides methods to convert between \Clock
instances and common time formats, such as np.datetime64 and float seconds 
since epoch, as well as comparing \Clock instances.

\subsubsection{Simulator}
The \Simulator class is instantiated with a \Clock instance and an instance 
each of \Inverter, \Battery, \Grid, \PowerSource, \Load, and (optionally) \pythonname{Context} respectively.
Its \step method takes \stepticks and optional keyword or positional arguments
for each component's step methods, calls each component's step method and 
returns all step results and a number of aggregates, such as generated, 
charged, discharged, consumed, and purchased energy per step and since the beginning of the simulation in Wh, e.g.

\[
\begin{aligned}
E_{\pythonname{PS},\text{cum},t} & =\sum_{t'\leq t}P'_{\pythonname{PS},t}\cdot\Delta t,\\
P_{\pythonname{G},\text{req},\max,t} & =\max_{t'\leq t}P_{\pythonname{G},\text{req},t}.
\end{aligned}
\]

In addition it tracks maxima of voltages (V), and currents (A) at each electric 
contact, to allow verifying that no limits are exceeded.

\subsection{Dataset Models}\label{sec:dataset-models}

The simulator provides comprehensive implementations of all aforementioned ABCs that 
expose the OpenCEM dataset via classes such as \pythonname{BatteryDataset}, 
\pythonname{GridDataset}, and etc.  They are all provided in the \pythonname{opencem.dataset} 
package.
Each constructor takes a clock instance, the ID of the inverter, where $1$ is
the ID of the inverter connected to the project's workstation and $2$ is the ID
of the inverter connected to the project office's AC, and a Sqlite3 database 
connection, supplying the dataset.

Because simulation step ticks and irregular real-world measurement times do not
align exactly due to the latency of the inverter's Modbus interface, the returned 
values are linear interpolations.
All models in the \pythonname{dataset} package ignore their inputs beyond 
\stepticks because they return fixed measurements.
A simple usage example for the dataset models is given in 
the public repository. For the sake of brevity we omit some of the code here, but because it is illustrative for the access to the context records, we provide the full listing in the linked repository.

\begin{figure}
\includegraphics[width=1.0\columnwidth]{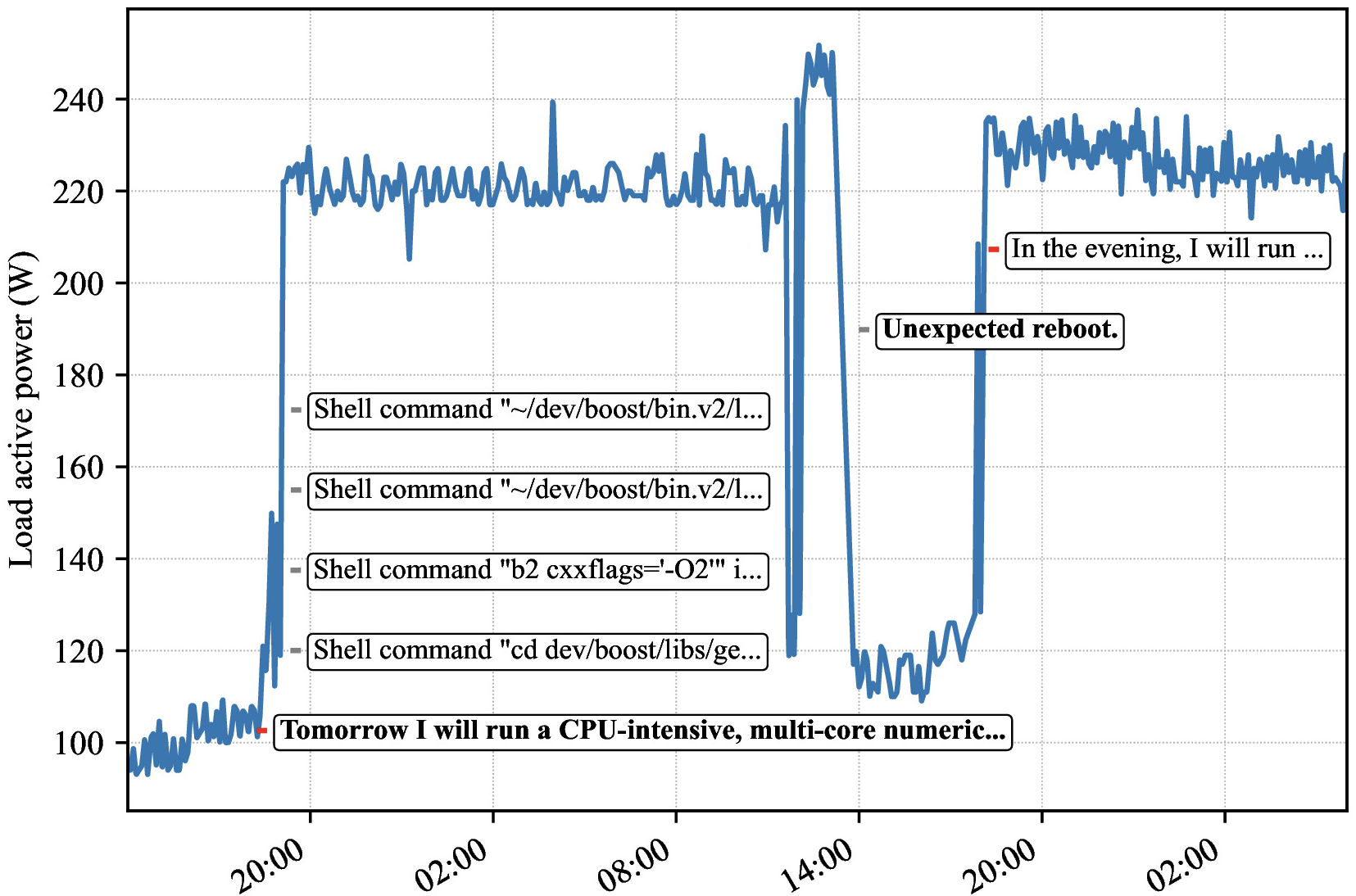}
  \vspace{-15pt}
  \caption{\textbf{Load Time Series} {\textnormal{from dataset models with selection of applicable context (2025-07-28). The example highlights a scenario in which the load estimate changes as new context becomes available, e.g. the unexpected reboot event falls into the time frame of the planned numerical stress test and becomes only available later. }}}
  \label{fig:load-time-series-with-context}
\end{figure}

In Figure~\ref{fig:load-time-series-with-context}, we show a combined visualization of time series load data overlayed with natural language descriptions of context events. On the date visualized here, one of the servers was used to run a CPU-intensive numerical robustness stress test, which was disrupted by an unplanned reboot. The examples highlights different modes of natural language context from manual entries by the user, shell command logs to system log entries. 

\subsection{Simulated Models}

While the dataset models implementing the simulator's interface provide convenient access to and computation with the OpenCEM dataset, the main goal is to simulate the result of alternative control strategies.
This requires the implementation of models that simulate how components would have behaved for control decisions.
For this purpose, the simulator package comes with additional models for the battery, grid, and inverter.
Additional models will be added in the future when more data is available for their verification.

\subsubsection{Linear Battery Model}
The simulator repository includes a \pythonname{linear} package, which provides
the model \pythonname{Bat\-te\-ry\-Li\-near}, implementing the \Battery ABC.
Its constructor takes a capacity $C$ in J, efficiencies $\eta_\text{charge}$, $\eta_\text{discharge}\in\left(0, 1\right]$ for charging and discharging 
as float, a fixed nominal voltage $U_\text{N}$ in V, an 
initial $\text{SOC}_{t_0}\in\left[0, 1\right]$, all with defaults that match the model in the on-campus installation. As internal state it initializes the current energy level $E_{t_0} = \text{SOC}_{t_0} \cdot C$.
The step function implements at each time $t$, for given \stepticks as $\Delta t$, a given battery mode out of IDLE, CHARGE, DISCHARGE, and an input current $I_{\text{bat},t}$ the following linear state update
$
    E_{t+\Delta t}=\max\{\min \{E_t+\Delta E, C \}, 0\},
$
with
\begin{align*}
\Delta E_{\pythonname{B},t}=\begin{cases}
0 & \text{if $\text{Mode}_t$=IDLE,}\\
\Delta t\cdot U_{\text{N}}\cdot I_{\pythonname{B},t}\cdot\eta_{\text{charge}} & \text{if $\text{Mode}_t$=CHARGING,}\\
-\Delta t\cdot U_{\text{N}}\cdot I_{\pythonname{B},t}\cdot\eta_{\text{discharge}}^{-1}  \! \! \! &\text{otherwise},
\end{cases}
\end{align*}
and returns as step result:
\[
\text{SOC}_{t +\Delta t} = \frac{E_{t + \Delta t}}{C}\in\left[0,1\right],
\]
as well as its constant nominal voltage, the unchanged input current, and discharge energy and capacity after clamping to the allowed range.

At the end of this section, we will verify empirically, that this approach is 
sufficiently adequate for scenarios in the dataset.

\subsubsection{Simple Inverter Model}
We provide a simple inverter model, \pythonname{In\-ver\-ter\-PV\-First}, which
balances power based on a simple priority order of PV-generated power, battery
reserves, and grid to match demand, and uses only surplus solar power to charge
the battery unless the configurable maximum SOC has been reached.
Besides efficiency values $\eta_{\pythonname{PV}\rightarrow\pythonname{B}}$, $\eta_{\pythonname{PV}\rightarrow\pythonname{L}}$, $\eta_{\pythonname{B}\rightarrow\pythonname{L}}$, $\text{SOC}_{\min}$, $\text{SOC}_{\max}$, the constructor accepts a constant
power $P_\pythonname{Inv}$ in W for the inverter's own power consumption.

\begin{figure}[!t]
\centering
\begin{tikzpicture}[
    scale=0.9,         
    transform shape,   
    font=\sffamily\scriptsize,
    node distance=0.8cm and 0.8cm,
    >={Latex[length=2mm, width=2mm]},
    block/.style={
        draw=black!70, 
        thick, 
        fill=white, 
        rectangle, 
        rounded corners=2pt, 
        align=center, 
        inner sep=5pt,
        drop shadow={opacity=0.1, shadow xshift=0.2ex, shadow yshift=-0.2ex}
    },
    layer/.style={
        draw=none,
        fill opacity=0.1,
        text opacity=1,
        inner sep=10pt, 
        rounded corners=5pt
    },
    layer_label/.style={
        font=\bfseries\tiny,
        anchor=north east, 
        inner sep=4pt
    },
    decision/.style={
        block,
        diamond,
        aspect=2.5,
        inner sep=1pt,
        text width=2.0cm
    },
    line/.style={draw=black!70, thick, ->}
]

    \node[block, fill=gray!10, text width=3.5cm] (input) {
        \textbf{Input}\\
        $\Delta t, P_{\mathrm{PS}}, P_{\mathrm{L,req}}, \text{SOC}_t$
    };

    \node[block, below=1.0cm of input, text width=4.0cm] (pv_calc) {
        \textbf{1. PV Allocation}\\
        $P_{\mathrm{net}} \leftarrow P_{\mathrm{PS}} - P_{\mathrm{L,req}}$
    };

    \node[decision, below=1.2cm of pv_calc] (check) {
        $P_{\mathrm{net}} \ge 0$?
    };
    
    \node[block, right=0.8cm of check, text width=2.0cm] (chg) {
        \textbf{Charge}\\
        $\min(P_{\mathrm{net}}, P_{\mathrm{max}})$
    };
    
    \node[block, left=0.8cm of check, text width=2.0cm] (dis) {
        \textbf{Discharge}\\
        $\max(P_{\mathrm{net}}, -P_{\mathrm{max}})$
    };

    \node[block, below=1.2cm of check, text width=4.0cm] (grid) {
        \textbf{3. Grid Import}\\
        $P_{\mathrm{G,req}} \leftarrow \text{Remaining Deficit}$
    };

    \node[block, fill=gray!10, below=1.0cm of grid, text width=3.5cm] (output) {
        \textbf{Output \& Update}\\
        $P_{\mathrm{G}}, I_{\mathrm{B}}, \text{Mode}, SOC_{t+1}$
    };

    \begin{scope}[on background layer]
        \node[layer, fill=blue, fit=(pv_calc), label={[layer_label, text=blue!60!black]north east:PRIORITY 1: PV}] (l1) {};
        
        \node[layer, fill=orange, fit=(check) (chg) (dis), label={[layer_label, text=orange!60!black]north east:PRIORITY 2: BATTERY}] (l2) {};
        
        \node[layer, fill=red, fit=(grid), label={[layer_label, text=red!60!black]north east:PRIORITY 3: GRID}] (l3) {};
    \end{scope}

    \draw[line] (input) -- (l1.north);
    \draw[line] (pv_calc) -- (check);
    
    \draw[line] (check.east) -- node[font=\tiny\bfseries, above] {YES} (chg.west);
    \draw[line] (check.west) -- node[font=\tiny\bfseries, above] {NO} (dis.east);
    
    \draw[line] (chg.south) |- ($(grid.north) + (0,0.3)$) -- (grid.north);
    \draw[line] (dis.south) |- ($(grid.north) + (0,0.3)$) -- (grid.north);
    
    \draw[line] (grid) -- (output);

\end{tikzpicture}
\vspace{-15pt}
\caption{\textbf{Hierarchical Control Logic.} The controller operates as a priority stack: (1) \textbf{PV Generation} is allocated first. (2) \textbf{Battery Storage} buffers any surplus or deficit. (3) \textbf{Grid Import} is used only as a last resort.}
\label{fig:inverter-step}
\end{figure}
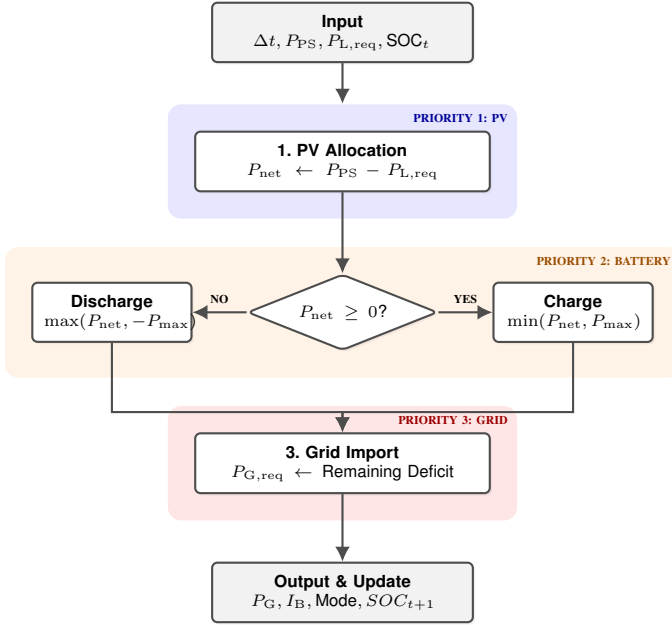

This closely matches the inverter's default configuration and can be used to 
verify the simulator's accuracy by comparing results to the dataset model.
Besides the default inputs, this model's step function accepts an optional power value $P_{\pythonname{G}\rightarrow\pythonname{B}}$ in W to allow external control decisions in charging the battery from grid power.
The dynamics are shown in Figure~\ref{fig:inverter-step}, which is limited to the case of $P_{\pythonname{G}\rightarrow\pythonname{B}} = 0$.

\subsubsection{Grid with Price Schedule}
We provide the model \pythonname{Grid\-Priced} which optionally accepts limits 
for active and apparent power in W and VA, respectively, and a schedule of 
electricity prices in cost units per kWh.
This is an example of how the simulator can be extended with a model that 
produces natural cost outputs for which optimizing policies can be tested or which can be used for, e.g., RL training in a Gymnasium environment.
Its step function returns results that extend \pythonname{Grid\-Step\-Re\-sult}
with a float field \pythonname{cost} and a boolean field 
\pythonname{vi\-o\-la\-tion}.

\subsubsection{Example and Validation}
In the linked repository, we provide usage demonstrations of the 
above models. Comparing the SOC of simulated linear models with the SOC curve 
previously obtained from the dataset, we find that the simulation closely 
matches the actual dataset over suitable periods.

\section{Simulation Scenarios and Applications}
\label{sec:simulation-examples}
In this section, we will provide two examples for using the dataset and simulator for evaluating context-aware prediction and control algorithms. Subsection~\ref{sec:prediction-example} shows how using natural language context improves power demand predictions over just using numerical metadata or no context. Subsection~\ref{sec:control-example} leverages this prediction approach for a battery management control scenario.

\subsection{Dateset Example for Context-Aware Prediction}
\label{sec:prediction-example}
For our first example, we demonstrate the benefit of using multi-modal context for power demand prediction. Sample data points for this can be conveniently obtained using the provided dataset models and database, as shown in the linked repository.
\begin{figure}[h]\includegraphics[trim={0.6cm 0.4cm 0.3cm 0.3cm}, clip, width=1.0\columnwidth]{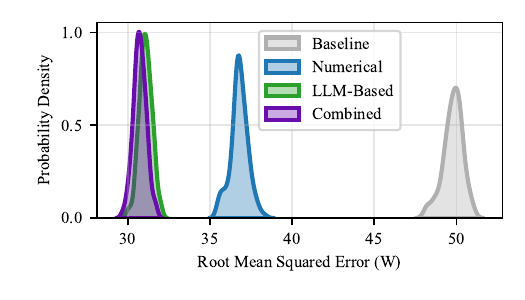}
  \vspace{-20pt}
  \caption{\textbf{Distribution of RMSE} in Watts for various context-aware power demand prediction models {\textnormal{trained on the dataset from 2025-10 to 2025-12. }}}
  \label{fig:context-aware-prediction-results}
\end{figure}
The multi-model context entries are provided as JSON dictionaries. For this experiment we predict the power demand for readings from October to December 2025. During this time, a large number of small CPU- and GPU-intensive jobs was run across both connected servers, yielding a total of 530 distinct context records. We compare prediction models that use 1. no context (as a baseline), 2. numerical context only such as the number of files processed in compilation jobs, the number of active CPU cores, or the number of model parameters in model fitting jobs, 3. natural language context which were transformed to numerical features by having the job effort estimated by a state of the art LLM (here GPT-5.2), and 4. a combination of numerical and natural language context.

\begin{figure}
  \includegraphics[trim={0.2cm 0.2cm 0.2cm 0.2cm}, clip, width=0.95\columnwidth]{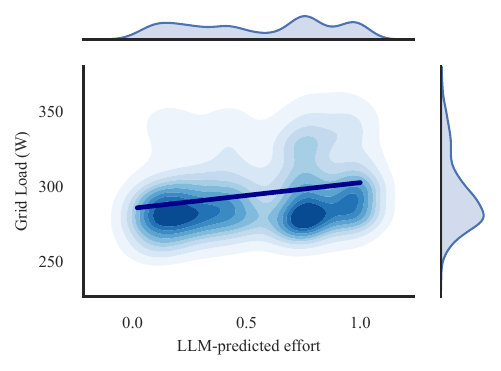}
  \vspace{-10pt}
  \caption{\textbf{Joint Distribution} of power demand and LLM-predicted effort for CPU task {\textnormal{based on the dataset filtered for records in which a CPU-intensive task was run only on one of the machines and no concurrent GPU jobs where running.}}}
  \label{fig:context-aware-prediction-effort-vs-load}
\end{figure}

The results are shown in Figure~\ref{fig:context-aware-prediction-results}. We find that there are significant improvements in prediction quality, when context information is used, and that using natural language context yields not only better results than numerical metadata but the numerical features seem to not provide any useful information beyond the effort estimate extracted from natural language in this example. The relationship between LLM-estimated effort and power consumption is shown for one task category in Figure~\ref{fig:context-aware-prediction-effort-vs-load}.
\vspace{-6pt}
\subsection{Simulation Example for Context-Aware Control}
\label{sec:control-example}

In the following we illustrate how the simulator can be used to
evaluate a context-aware control strategy.
A natural optimization problem for such a system with dynamic grid pricing is to decide when to charge the battery in advance during off-peak hours to avoid buying expensive energy for future load during peak hours.
\begin{figure}[h]
\centering
\begin{tikzpicture}[
    scale=0.8, 
    transform shape,
    font=\sffamily\footnotesize,
    >={Latex[length=2mm, width=2mm]},
    env_block/.style={
        draw=blockBorder, 
        thick, 
        fill=simColor,
        rectangle,
        rounded corners=2pt, 
        align=center, 
        minimum height=1.2cm, 
        minimum width=2.8cm,
        inner sep=4pt,
        drop shadow={opacity=0.15, shadow xshift=0.5ex, shadow yshift=-0.5ex}
    },
    agent_block/.style={
        draw=blockBorder, 
        thick, 
        fill=white,
        rectangle,
        rounded corners=2pt, 
        align=center, 
        minimum height=1.0cm, 
        minimum width=2.5cm,
        inner sep=4pt,
        drop shadow={opacity=0.1, shadow xshift=0.5ex, shadow yshift=-0.5ex}
    },
    arrow_label/.style={
        font=\scriptsize\color{blockBorder}, 
        inner sep=2pt,
        align=center
    }
]


    \node[env_block] (sim) {
        \textbf{OpenCEM Simulator}\\
        \textit{(Environment)}\\
        \scriptsize Generates $x_t, \text{Logs}$
    };

    \node[agent_block, right=2.5cm of sim] (llm) {
        \textbf{Semantic}\\
        \textbf{Interpreter}\\
        \textit{(LLM)}
    };

    \node[agent_block, below=1.2cm of llm] (pred) {
        \textbf{Load}\\
        \textbf{Forecaster}\\
        \textit{(Regression)}
    };

    \node[agent_block, below=1.2cm of sim] (mpc) {
        \textbf{MPC}\\
        \textbf{Controller}\\
        \textit{(Optimization)}
    };


    \draw[->, thick, blockBorder] (sim.east) -- node[arrow_label, midway, above] {Unstructured\\Context (Logs)} (llm.west);

    \draw[->, thick, blockBorder] (llm.south) -- node[arrow_label, midway, right] {Effort\\Features} (pred.north);

    \draw[->, thick, blockBorder] (pred.west) -- node[arrow_label, midway, above] {Predicted Load\\$\hat{P}_L$} (mpc.east);

    \draw[->, thick, blockBorder] (mpc.north) -- node[arrow_label, midway, left] {Grid Import\\$u^*_t$} (sim.south);

    \draw[->, thick, blockBorder] (sim.west) -- ++(-0.8, 0) |- node[arrow_label, near start, left] {State Feedback\\$x_t$ (SOC, etc.)} (mpc.west);

    \begin{scope}[on background layer]
        \node[draw=orange!60, dashed, thick, fill=orange!5, rounded corners=6pt, fit=(llm) (pred) (mpc), inner sep=12pt] (agent_group) {};
        \node[anchor=north east, font=\bfseries\scriptsize\color{orange!70!black}, inner sep=4pt] at (agent_group.north east) {Application Example: Context-Aware Agent};
    \end{scope}
\end{tikzpicture}
  \vspace{-15pt}
\caption{\textbf{Validation Use Case}: Context-Aware Control Loop. The figure illustrates the interaction between the \textbf{OpenCEM Simulator} and an example \textbf{LLM-based Control Agent}. Arrows represent data flow, with labels positioned alongside to ensure visibility.}
\label{fig:context-based-optimisation-algorithm}
\end{figure}
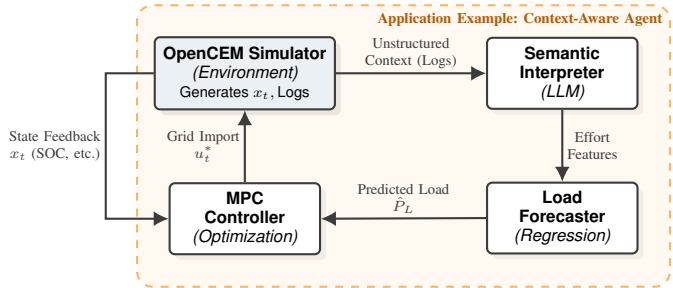

Ignoring inefficiencies for simplicity of presentation, we can formulate the problem as follows: 
\begin{subequations}
\label{eq:cost_optimization}
\begin{align}
\min_{\{P^\text{control}_{\mathrm{G},\text{req},t}\}_{t=t_0}^{T-1}}
&\quad \sum_{t=t_0}^{T-1} \pi_t \frac{\Delta t}{3.6\times 10^6}\, P^\text{control}_{\mathrm G, \text{req},t}
\\
\text{s.t.}\quad
& 0 \le P_{\mathrm G, \text{req},t},
\\
& P_{B,t} = P_{PS,t} + P_{G,t} - P_{L,\mathrm{req},t},
\\
& SOC_{t+1} = SOC_t + \frac{\Delta t}{C}\, P_{B,t},
\\
& SOC_{\min} \le SOC_t \le SOC_{\max},
\\
& SOC_{t_0} \text{ given},
\qquad t=t_0,\dots,T-1,
\end{align}
\end{subequations}
where $\pi_t$ is the price of electricity for a given schedule.

For given future load, and power generation time series up to some time horizon as well as given day-ahead pricing, the minimal cost solution, in terms of amount of energy to buy from the grid at each time step, can be obtained with standard optimization techniques.
In real-life applications, predictions of load can change throughout the time horizon.
Examples of this are represented in the OpenCEM dataset, see e.g. Figure~\ref{fig:load-time-series-with-context}: The robustness test was scheduled a day ahead for 24 hours, but load dropped prematurely due to an unexpected reboot.
For the online problem with changing predictions motivated by this, we can at each time step obtain a prediction for some horizon, compute the optimal action, use the action in the first time step and discard the remainder.
This approach allows to account for changes in prediction and context.

In the following Figure~\ref{fig:context-based-optimisation-algorithm}, we illustrate how to run an experiment with this strategy using the simulator introduced in this work and an online learning algorithm based on the ideas presented in~\cite{instructmpc}, using a combination of context-based prediction and classical optimization algorithms to minimize cost for a given power demand, electricity cost, PV generation, and battery constraints.
\begin{figure}[h]  \includegraphics[width=1.0\columnwidth]{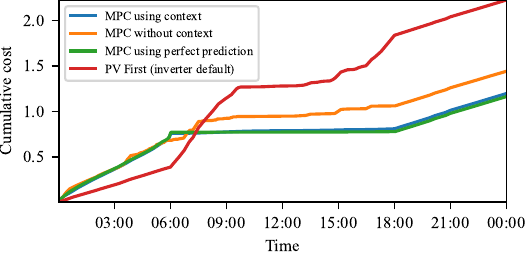}
  \vspace{-20pt}
  \caption{Running cost in control experiment {\textnormal{for the day 2025-12-25 of the dataset, using the inverter default strategy (green), predictions based on prior consumption (orange), predictions based on LLM-processed context (blue) and perfect predictions as a benchmark (red). We find that using context-based yields near-optimal results. The default policy has lower initial cost but higher final cost because it fails to buy power during cheap hours to meet later demand. }}}
  \label{fig:context-based-optimisation-results}
\end{figure}
Using our dataset, we find that MPC with predictions based on features extracted from natural language context yields near-optimal and significant improvements over the inverters default strategy (PV first), and context-less predictions based on prior demand, see Figure~\ref{fig:context-based-optimisation-results} for different strategies on a single day, and Figure~\ref{fig:cost-heatmap} for the cost-savings using our proposed strategy over the inverter default settings across multiple days.

\begin{figure*}\centering\includegraphics[width=1.09\textwidth]{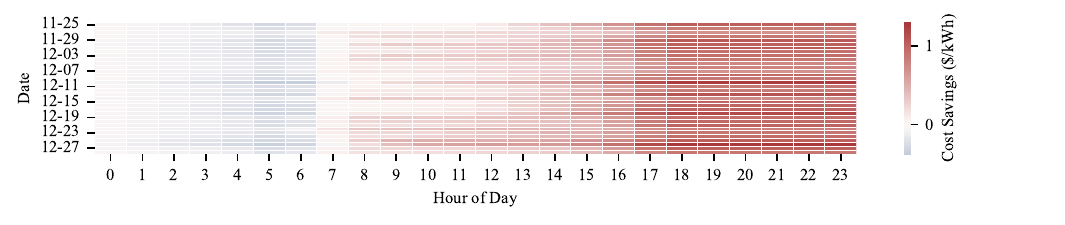}
  \vspace{-25pt}
  \caption{Running cost savings in the control experiment {\textnormal{for the days 2025-11-25 to 2025-12-30 of the dataset, using the proposed strategy with predictions based on natural language context over the inverter default strategy. The color indicates the cumulative cost savings by hour of day. We find that using context-based yields near-optimal results. We find that the savings shown in Figure~\ref{fig:context-based-optimisation-results} are consistent across multiple days. }}}
  \label{fig:cost-heatmap}
\end{figure*}

\section{Conclusion and Outlook}
In this work, we introduced the OpenCEM simulator, the first open-source digital twin designed to explicitly integrate contextual information into renewable energy management. Unlike existing simulators that are limited to physical dynamics, OpenCEM combines physical simulation, with a rich dataset and natural language context, enabling the development and validation of context-aware control strategies in a microgrid setting.

The validation against real-world data shows that the provided simulator models can faithfully reproduce the core system behaviors and dynamics, while its support for user-generated and automatically extracted context records enables a new class of control and forecasting methods powered by large language models and other AI approaches. By making the dataset and simulator openly available, we aim to lower the barrier for researchers and support innovation in context-aware control algorithms for sustainable  energy systems. We demonstrated how the simulator's modular and extensible API provides a flexible framework for validating physical models, testing new control strategies, and explore the relationship of human- and machine-generated unstructured context and load in a real-world installation.

For upcoming work, we plan to extend the simulator in several directions. First, we will release longer time series covering a wider range of operational conditions and workloads to enhance the robustness of models trained on the OpenCEM dataset. Second, we will expand the physical installation with additional servers to enrich the diversity of load patterns and contextual information. Third, more sophisticated surrogate models for lithium-ion batteries and PV generation will be added, enabling higher-fidelity simulations that capture nonlinearities and degradation effects. Finally, we will provide integration with RL frameworks to provide a benchmark environment for optimization, reinforcement learning, and in-context reasoning.

By addressing both the data and simulation gaps in contextual energy research, OpenCEM provides a step toward intelligent, context-aware renewable energy systems. We hope that this platform will not only encourage new research but also help with broader adoption of context-aware approaches to decarbonizing the grid.

\bibliographystyle{IEEEtran}
\bibliography{references}

\begin{thebibliography}{10}
\providecommand{\url}[1]{#1}
\csname url@samestyle\endcsname
\providecommand{\newblock}{\relax}
\providecommand{\bibinfo}[2]{#2}
\providecommand{\BIBentrySTDinterwordspacing}{\spaceskip=0pt\relax}
\providecommand{\BIBentryALTinterwordstretchfactor}{4}
\providecommand{\BIBentryALTinterwordspacing}{\spaceskip=\fontdimen2\font plus
\BIBentryALTinterwordstretchfactor\fontdimen3\font minus \fontdimen4\font\relax}
\providecommand{\BIBforeignlanguage}[2]{{%
\expandafter\ifx\csname l@#1\endcsname\relax
\typeout{** WARNING: IEEEtran.bst: No hyphenation pattern has been}%
\typeout{** loaded for the language `#1'. Using the pattern for}%
\typeout{** the default language instead.}%
\else
\language=\csname l@#1\endcsname
\fi
#2}}
\providecommand{\BIBdecl}{\relax}
\BIBdecl

\bibitem{impram2020challenges}
S.~Impram, S.~V. Nese, and B.~Oral, ``Challenges of renewable energy penetration on power system flexibility: A survey,'' \emph{Energy strategy reviews}, vol.~31, p. 100539, 2020.

\bibitem{dong2024surveyincontextlearning}
\BIBentryALTinterwordspacing
Q.~Dong, L.~Li, D.~Dai, C.~Zheng, J.~Ma, R.~Li, H.~Xia, J.~Xu, Z.~Wu, T.~Liu, B.~Chang, X.~Sun, L.~Li, and Z.~Sui, ``A survey on in-context learning,'' 2024. [Online]. Available: \url{https://arxiv.org/abs/2301.00234}
\BIBentrySTDinterwordspacing

\bibitem{instructmpc}
R.~Wu, J.~Ai, and T.~Li, ``Instructmpc: A human-llm-in-the-loop framework for context-aware control,'' in \emph{2025 IEEE 64th Conference on Decision and Control (CDC)}, Dec 2025, pp. 172--179.

\bibitem{dataset_microgrid_mesa_del_sol}
A.~Bashir, C.~Leap, A.~Blumenthal, T.~Estrada, A.~Bidram, M.~Martinez-Ramon, and M.~Abdullah, ``Power, voltage, frequency and temperature dataset from mesa del sol microgrid,'' Aug. 2023.

\bibitem{dataset_microgrid_nanogreen_japan}
K.~Vink, E.~Ankyu, and M.~Koyama, ``\BIBforeignlanguage{en}{Multiyear microgrid data from a research building in tsukuba, japan},'' \emph{\BIBforeignlanguage{en}{Sci. Data}}, vol.~6, no.~1, p. 190020, Feb. 2019.

\bibitem{dataset_microgrid_rye}
P.~Aaslid, ``Rye microgrid load and generation data, and meteorological forecasts,'' 2021.

\bibitem{fernandez2019power}
A.~Fern{\'a}ndez-Guillam{\'o}n, E.~G{\'o}mez-L{\'a}zaro, E.~Muljadi, and {\'A}.~Molina-Garc{\'\i}a, ``Power systems with high renewable energy sources: A review of inertia and frequency control strategies over time,'' \emph{Renewable and Sustainable Energy Reviews}, vol. 115, p. 109369, 2019.

\bibitem{moeini2025surveyincontextreinforcementlearning}
\BIBentryALTinterwordspacing
A.~Moeini, J.~Wang, J.~Beck, E.~Blaser, S.~Whiteson, R.~Chandra, and S.~Zhang, ``A survey of in-context reinforcement learning,'' 2025. [Online]. Available: \url{https://arxiv.org/abs/2502.07978}
\BIBentrySTDinterwordspacing

\bibitem{opencem_demo}
\BIBentryALTinterwordspacing
Y.~Lu, T.~S. Bartels, R.~Wu, F.~Xia, X.~Wang, Y.~Wu, H.~Yang, and T.~Li, ``Open in-context energy management platform,'' in \emph{Proceedings of the 16th ACM International Conference on Future and Sustainable Energy Systems}, ser. E-Energy '25.\hskip 1em plus 0.5em minus 0.4em\relax New York, NY, USA: Association for Computing Machinery, 2025, p. 985–986. [Online]. Available: \url{https://doi.org/10.1145/3679240.3734678}
\BIBentrySTDinterwordspacing

\bibitem{5491276}
R.~D. Zimmerman, C.~E. Murillo-Sánchez, and R.~J. Thomas, ``Matpower: Steady-state operations, planning, and analysis tools for power systems research and education,'' \emph{IEEE Transactions on Power Systems}, vol.~26, no.~1, pp. 12--19, 2011.

\bibitem{cui2020hybridsymbolicnumericframeworkpower}
\BIBentryALTinterwordspacing
H.~Cui, F.~Li, and K.~Tomsovic, ``Hybrid symbolic-numeric framework for power system modeling and analysis,'' 2020. [Online]. Available: \url{https://arxiv.org/abs/2002.09455}
\BIBentrySTDinterwordspacing

\bibitem{lara2024powersimulationsdynamicsjlopensource}
\BIBentryALTinterwordspacing
J.~D. Lara, R.~Henriquez-Auba, M.~Bossart, D.~S. Callaway, and C.~Barrows, ``Powersimulationsdynamics.jl -- an open source modeling package for modern power systems with inverter-based resources,'' 2024. [Online]. Available: \url{https://arxiv.org/abs/2308.02921}
\BIBentrySTDinterwordspacing

\bibitem{Anderson2023}
\BIBentryALTinterwordspacing
K.~S. Anderson, C.~W. Hansen, W.~F. Holmgren, A.~R. Jensen, M.~A. Mikofski, and A.~Driesse, ``pvlib python: 2023 project update,'' \emph{Journal of Open Source Software}, vol.~8, no.~92, p. 5994, 2023. [Online]. Available: \url{https://doi.org/10.21105/joss.05994}
\BIBentrySTDinterwordspacing

\bibitem{8909765}
Z.~J. Lee, D.~Johansson, and S.~H. Low, ``Acn-sim: An open-source simulator for data-driven electric vehicle charging research,'' in \emph{2019 IEEE International Conference on Communications, Control, and Computing Technologies for Smart Grids (SmartGridComm)}, 2019, pp. 1--6.

\bibitem{Orfanoudakis_2025}
\BIBentryALTinterwordspacing
S.~Orfanoudakis, C.~Diaz-Londono, Y.~Emre~Yılmaz, P.~Palensky, and P.~P. Vergara, ``Ev2gym: A flexible v2g simulator for ev smart charging research and benchmarking,'' \emph{IEEE Transactions on Intelligent Transportation Systems}, vol.~26, no.~2, p. 2410–2421, Feb. 2025. [Online]. Available: \url{http://dx.doi.org/10.1109/TITS.2024.3510945}
\BIBentrySTDinterwordspacing

\bibitem{SCHWARZ2021100839}
\BIBentryALTinterwordspacing
S.~Schwarz, S.~A. Uerlich, and A.~Monti, ``pycity\_scheduling—a python framework for the development and assessment of optimisation-based power scheduling algorithms for multi-energy systems in city districts,'' \emph{SoftwareX}, vol.~16, p. 100839, 2021. [Online]. Available: \url{https://www.sciencedirect.com/science/article/pii/S2352711021001230}
\BIBentrySTDinterwordspacing

\bibitem{7750303}
J.~S. Stein, W.~F. Holmgren, J.~Forbess, and C.~W. Hansen, ``Pvlib: Open source photovoltaic performance modeling functions for matlab and python,'' in \emph{2016 IEEE 43rd Photovoltaic Specialists Conference (PVSC)}, 2016, pp. 3425--3430.

\bibitem{10.1145/3307772.3328313}
\BIBentryALTinterwordspacing
Z.~J. Lee, T.~Li, and S.~H. Low, ``Acn-data: Analysis and applications of an open ev charging dataset,'' in \emph{Proceedings of the Tenth ACM International Conference on Future Energy Systems}, ser. e-Energy '19.\hskip 1em plus 0.5em minus 0.4em\relax New York, NY, USA: Association for Computing Machinery, 2019, p. 139–149. [Online]. Available: \url{https://doi.org/10.1145/3307772.3328313}
\BIBentrySTDinterwordspacing

\bibitem{brockman2016openaigym}
\BIBentryALTinterwordspacing
G.~Brockman, V.~Cheung, L.~Pettersson, J.~Schneider, J.~Schulman, J.~Tang, and W.~Zaremba, ``Openai gym,'' 2016. [Online]. Available: \url{https://arxiv.org/abs/1606.01540}
\BIBentrySTDinterwordspacing

\bibitem{modbus2012}
\BIBentryALTinterwordspacing
{Modbus Organization, Inc.}, \emph{MODBUS Application Protocol Specification V1.1b3}, Apr. 2012, available online. [Online]. Available: \url{https://modbus.org/specs.php}
\BIBentrySTDinterwordspacing

\end{thebibliography}

\end{document}